\documentclass[conference,letterpaper]{IEEEtran}

\usepackage{cite}      

\usepackage{stfloats}  

\usepackage[latin1]{inputenc}
\usepackage[T1]{fontenc}
\usepackage{textcomp}
\usepackage{subfigure}
\usepackage{amssymb,amsmath,xspace}
\providecommand{\href}[2]{#2} 
\providecommand{\hypersetup}[1]{}\providecommand{\url}[1]{#1}
\usepackage[dvips]{graphicx} 
\usepackage[hypertex]{hyperref} 
\newcommand{\ie}{, {i.e.},\xspace}
\newcommand{\eg}{, {e.g.},\xspace}

\newcommand{\dist}{\operatorname{dist}}
\begin{document}

\title{Trust for Location-based Authorisation}

\author{\authorblockN{Andreas U.\ Schmidt\authorrefmark{1},}
\authorblockN{Nicolai Kuntze\authorrefmark{1}, J\"org Abendroth\authorrefmark{2},}
\authorblockA{\authorrefmark{1}Fraunhofer Institute for Secure Information Technology SIT\\
Rheinstra\ss e 75, 64295 Darmstadt, Germany\\
Email: \{andreas.schmidt,nicolai.kuntze\}@sit.fraunhofer.de}
\authorblockA{\authorrefmark{2}Nokia Siemens Networks GmbH \& Co. KG\\
Otto-Hahn-Ring 6, 81737 Munich, Germany\\
Email: joerg.abendroth@nsn.com}}


%


\maketitle

\begin{abstract}
We propose a concept for authorisation using the location of a mobile device
and the enforcement of location-based policies.
Mobile devices enhanced by Trusted Computing capabilities operate
an autonomous and secure location trigger and policy enforcement entity.
Location determination is two-tiered, integrating cell-based triggering at
handover with precision location measurement by the device.
\end{abstract}


%
\IEEEpeerreviewmaketitle
\section{Introduction}\label{sec:introduction}
Ambient technologies propose to enrich the mobile environment with different
computer devices and networks. One use case is to utilise a user device 
for remote control of other systems --- to
 automatically turn on the light when entering
a room, or to enable access to the emergency controls of a power plant
while repairing the pipes. Often, a natural requirement is to restrict
possibilities to authorised personnel in a certain location.
Vice versa sometimes one wants to restrict the functionality of
devices based on its location\eg disable the
camera while entering a sensitive area. All these scenarios require
location based authorisation and sometimes access control,
a subject which has attracted some attention in the literature,
cf.~\cite{Anis07} and references therein.

The implementation of location based authorisation can be
based on various methods for location determination\eg 
triangulation (e.g.\ strength of the device signal at six
base stations~\cite{Woel02,WHZ+02}), 
calculating the round trip time~\cite{VNJ02}, or
deploying a GPS module locally to the client. The first method
lacks the precision, and the second requires 
new sensors to be installed in the location-based authorisation
enabled areas. The last has not been deployed because relying on the
client side for location information is deemed insecure. 
Event though  hybrid approaches like combination of triangulation and GPS
exist, the latter argument applies to them.
In conclusion current approaches are generally unfit for authorisation 
and access control based on location with scalable
strength of enforcement.
We propose a
solution, that combines three technologies. 
First a network-side 
location trigger, which only has cell granularity,
and second a device-side enforcing agent using de-central (e.g.\ GPS)
or co-operative (e.g.\ A-GPS) location methods and concurrently
enforces the pertinent policies.
Third security is established by applying Trusted Computing concepts
of Attestation.
Section~\ref{sec:tlta-concept} presents the proposed concept
on the level of functional building blocks, architecture, and 
operational scenario. Section~\ref{sec:protocols-procedures}
defines the required communication protocols and their integration
into pcs networks. 
Security and efficiency of the concept and application scenarios are
discussed in Section~\ref{sec:discussion}.
Section~\ref{sec:conclusion} concludes the paper.
\section{TLTA Concept}\label{sec:tlta-concept}
In order to fell authorisation decisions based on device location,
an entity called \textit{location trigger} must have access to location information
and be able to provide this information to other entities enforcing according policies,
either on the device or on the part of some service.
The concept of an Trusted Location Trigger Authorisation (TLTA) we propose in the
following, rests on these key ideas.
First, the location trigger in TLTA operates on two separate levels: 
i) Handover-based localisation within a network cell and 
ii) GPS-based (here always meaning either full, device operated GPS or A-GPS)
localisation within an area circumscribed by a perimeter of cells.
Second, a trusted entity on the device, called the Location Trigger Enforcer (LTE),
embodying the location trigger functionality and enforcing, possibly in co-operation
with a network-side counterpart, authorisation policies.
The trust in the LTE rests on security properties of the device, which we assume to
be a trusted computing (TC) platform for the purpose of location-based authorisation.
The latter notion is, for our subject, specified by the standards of the
Trusted Computing Group's (TCG) Mobile Phone Working Group (MPWG, see~\cite{MPWGRefArch}).
The necessary architectural principals of the MPWG for TLTA are explained in
Section~\ref{sec:trusted-mts}. 
It should be noted however, that the TLTA concept is applicable resting 
on any other trusted platform architecture which fulfils the core requirements in
the namely section.

In the following, we will take an agnostic stance with respect to 
the actual restrictions that are to be enforced and just speak of `the policies'
to be enforced.
The next subsection sketches the architecture of a TLTA system and security-relevant
assumptions.
The state sequence of a device operating under the TLTA scheme is
developed in Section~\ref{sec:mobility-process}.
\subsection{Entities and their functions}\label{sec:entit-their-funct}
We here choose the Long Term Evolution of 3G Networks
specified by the 3GPP  standardisation (3GPPLTE, see~\cite{3GPPLTE})
as the technical framework for the presentation of TLTA.
Within it, various improvements over common 2G and 3G network architectures
are specified. One of the aims is to simplify the overall architecture.
This flat, all-IP based infrastructure is referred to as System Architecture
Evolution (SAE). Meeting technical and performance
requirements arising from an analysis of the current state of mobile 
technology resulted in a reduction of the number of network nodes involved in data processing and
transport. A flatter network architecture leads to improved data latency (the
transmission delay between the transmitter sending data and the time
of reception) and better support of delay-sensitive, interactive and real-time
communications. 

\begin{figure}[!h]
\centerline{\subfigure{\resizebox{0.25\textwidth}{!}{\includegraphics{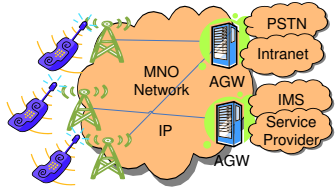}}}} 
\caption{SAE model and its basic components}
  \label{fig:SAE}
\end{figure}
The SAE model is shown in Figure \ref{fig:SAE} and consists 
of two types of network elements. Access of user devices, called Mobile Terminals (MT)
is facilitated by enhanced base stations, referred to as enhanced Node-Bs (eNB).
They provide the air interface and performs the resource management thereof.
The second class of elements are the Access Gate Ways (AGW) which implement  
management functionalities and act as data gateways for user packages.
The AGW establishes connections to Service Providers.
Within this core network model the MNO can offer different services, for instance A-GPS.

\begin{figure}[!t]
\centerline{\subfigure{\resizebox{0.3\textwidth}{!}{\includegraphics{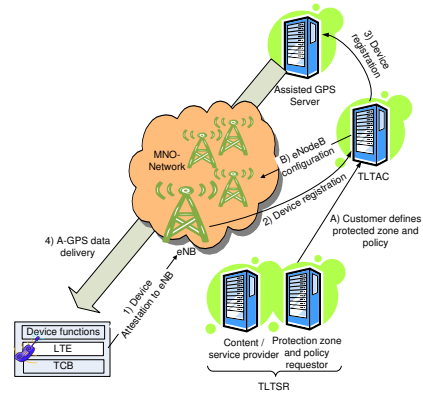}}}} 
\caption{Interaction of an existing MNO network and the TLTA components}
  \label{fig:TLTA_szenario}
\end{figure}

Figure \ref{fig:TLTA_szenario} depicts the interaction between the basic SAE
architecture (shown as an MNO cloud) and the components added by TLTA.
The customer who requests the enforcement of certain policies on devices in
a geographic area, called the  \textit{protected zone},
is termed TLTAC service requester (TLTSR).
The TLTSR handles for instance the provisioning of content to devices.  
TLTA introduces on network operator side the Trusted Location 
Trigger Authorisation Centre (TLTAC) which establishes 
a pre-defined geographic area within the area covered by the MNO network. 
The TLTAC receives (i) the physical geometry of the
protected zone and (ii) the desired policies to be enforced within this zone
(A in the figure). 
Based on these data the TLTAC determines  
configures certain eNBs of the MNO in Step B. 
After the setup TLTAC's main 
task is to authenticate the end users devices and to maintain a list of them
for authorisation purposes.

A Location Trigger Enforcer (LTE) is integrated in the end users MT
which is the device-side counterpart of the TLTAC 
(The abbreviation LTE used in this paper should 
not be confounded with Long Term Evolution in the context of the 3GPP). 
LTE (i) receives and holds the 
policies from TLTAC, (ii) receives and holds the protection zone geometry,
(iii) determines entry and exit events to the protection zone by some high-precision location
mechanism\eg A-GPS, and 
(iv) enforces the policies on the device.

After this setup,  MTs entering the protected zone 
may acquire certain services. The access to these services is
controlled by the LTE. The basic scheme is that the
device which enters the protected zone performs an attestation (see below) of its 
internal state and configuration to the MNO (Step~1 in Figure~\ref{fig:TLTA_szenario}). 
If this attestation succeeds the device is registered at the TLTAC (Step 2)
and also registered with the A-GPS service, or any other location providing service 
which is used to determine the 
accurate location of the device (Step 3). These registration steps are mainly
required to protect the location service against fraud and to enable certain 
payment methods and schemes.
\subsection{Trusted MTs}\label{sec:trusted-mts}
The TLTAC and LTE together form a Policy Decision Point (PDP) / Policy Enforcement Point (PEP)
related model based on a local enforcement of policies. 
From the perspective of the content provider the trustworthiness of the local LTE 
must be testified and sufficiently strong. Therefore, TLTA requires that the MT 
constitutes a trusted platform resting on certain capabilities constituting a
Trusted Computing Base (TCB),
which for the purpose of this report comprise:
\subsubsection{Secure boot} 
The boot process is the time span from the power on to the stage when the user is able 
to interact with the device. Within this time span, secure boot 
ensures that fundamental capabilities are in place 
which are critical for the reliability of the system, comprising in particular
 means for resource isolation and management.
During secure boot, as defined in~\cite{MPWGRefArch}, each component is measured in order of execution. 
These measurement values
are then verified by a local verification agent using so called Reference Integrity Measurement (RIM) 
digital certificates. 
If the values match then the processing continues and the started component is registered in
a log as well as in a corresponding system state register in a secure storage space. 
Otherwise the boot will fail
and the system switches into a failed state.
In case of a failed state the system may offer a Pristine Boot which
as a fallback. Pristine Boot defines a device state after
initial factory installation where the life cycle of a device begins. 
Another use case for 
the Pristine Boot is an update process to rebuild the 
required credentials (\eg RIM certificates) used during secure boot.
If the secure boot of an MT succeeds the system on top can rely on the
integrity of the system underneath. This includes all subsystems integrated in the
platform including the LTE. 
\begin{figure*}[!t]
\centerline{\subfigure{\resizebox{0.6\textwidth}{!}{\includegraphics{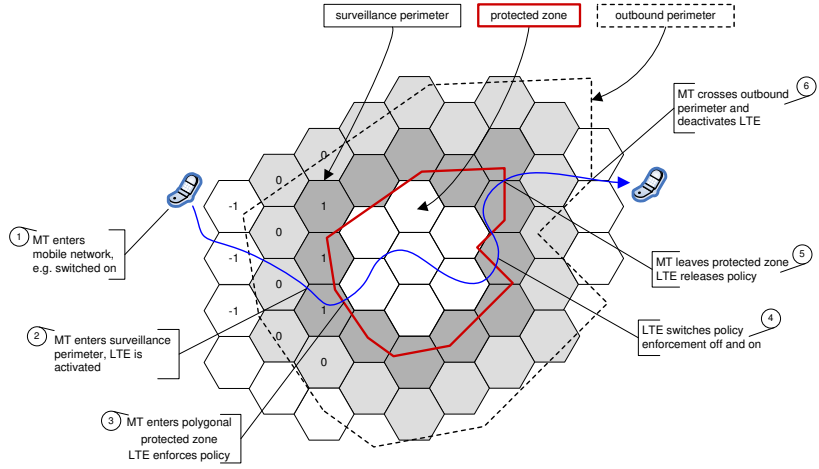}}}} 
\caption{Mobile station travelling through the zones and perimeters of a TLTA system.}
  \label{fig:TLTAC_zones_and_perimeters}
\end{figure*}
\subsubsection{Remote attestation}
A system which has performed a secure boot process on an appropriate platform
constitutes a trustworthy base for other operations. 
This is not useful without means to convey
this state in a reliable way to other entities.
This process is called attestation in the context of TC.
TC uses two concepts to implement this functionality: Platform Configuration Registers
(PCR) and Attestation Identities (AI). A PCR reflects the system configuration reported by 
the measurement entity and is protected within the Mobile Trusted Module (MTM) against tampering. 
PCR values are created by combining digest values representing the measured component. Each value 
added to the PCR is also recorded in a log. Log and the corresponding PCR value together
are testifying to the system state. To simplify the verification process, the MPWG 
has introduced a device side verifier. This verifier works as an agent
close with the measurement system and holds RIM certificates. 
An Attestation Identity is embodied in a credential, and usually certified by
a trusted third party\eg a Privacy CA, to attest that the owning MT contains
a live, unaltered MTM. Producing the AI credential together with PCR values
and measurement logs and signed by the AI private key to an external 
verifier, constitutes the attestation process proper.
The verifier, in the particular case at hand, then has assurance
that the MT operates an unaltered LTE on top of a set of trustworthy resources.
\subsection{Mobility process}\label{sec:mobility-process}
The \textit{protected zone}, denoted $pz$, 
is for simplicity taken as the area circumscribed by a closed
polygon in the plane which in turn is covered by cells of the mobile
network.
Now, assign the number $1$ to all cells on the boundary of the 
closed, connected set of cells covering the $pz$ and call this set
of cells $c_1$
The \textit{surveillance perimeter} ($sp$) is defined as outer boundary
of $c_1$.
Call $c_0$ the set of cells adjacent to $sp$, $c_{-1}$ the set of cells adjacent
to the outer boundary of $c_0$, and so on.
Finally let the \textit{outbound perimeter} $op$ be a closed polygon
such that $c_1\subset\overline{op}$ and $\dist(op,c_1)>0$.
This geometry is shown in  Figure~\ref{fig:TLTAC_zones_and_perimeters},
wherein, $op$ results from simply scaling $pz$.  

Before TLTA can operate according to a TLTSR request, steps of 
device preparation and node configuration must be executed.
For the latter, TLTAC transmits a request to the AGW which
configures the eNBs of  $c_0$ and $c_1$ to execute the trust-enhanced
handover described in Section~\ref{sec:trust-enhanc-hand} below, when
an MT crosses the surveillance perimeter inbound.
Device preparation is not required for trusted MTs already equipped with
an LTE, thus it reduces to LTE enrolment.
This can be done by secure software download,
which is a major use case for TC-enabled devices, at any stage
in the mobility process\eg when a device enters a $c_0$ cell.

\begin{figure}[!h]
\centerline{\subfigure{\resizebox{0.45\textwidth}{!}{\includegraphics{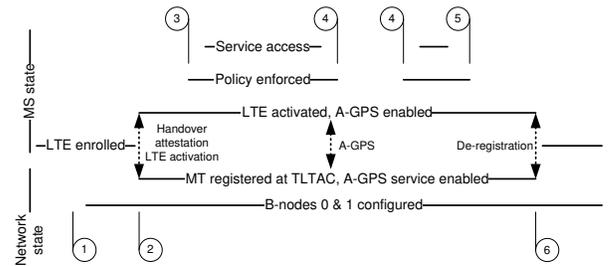}}}} 
\caption{Network and device states during the travel.}
  \label{fig:TLTAC_state_evolution}
\end{figure}
Figure~\ref{fig:TLTAC_zones_and_perimeters} shows the travel of a
 mobile device MT through the zones.
We describe this journey on a functional level, deferring details
of communication protocols and procedures to Section~\ref{sec:protocols-procedures}.
The states of MT and the mobile network relevant to TLTA are shown
in Figure~\ref{fig:TLTAC_state_evolution}.  
At the point marked $(1)$, MT enters the mobile network in a normal operational mode,
with deactivated LTE and A-GPS functions.
Upon crossing $sp$ at $(2)$, the handover between a cell of $c_0$ and the corresponding
cell of $c_1$ is augmented by the activation of the LTE and the A-GPS,
a remote attestation process, registration of MT with the TLTAC, and
download of the authorisation policy, and the geometry of $pz$ to MT.
From there on the device autonomously operates the location trigger to
detect $p$ using A-GPS and with the aid of an A-GPS support server
(possibly a function of TLTAC).
When MT enters $pz$ at $(3)$, LTE enforces the policy, for instance unlocks
a preassigned credential for service access.
During sojourn in $pz$, MT follows the policy and can for instance use a 
LBS whenever the user desires.
At some points $(4)$, MT may leave and re-enter $pz$ and LTE switches the
enforcement of the policy off and on.
LTE finally releases policy enforcement at point $5$, but is deactivated only at
crossing of $op$.
At $(6)$, LTE also actively de-registers MT with TLTAC.
The triggering of location is cell-based only between points $(1)$ and $(2)$ and
after $(6)$, otherwise it relies on (assisted) GPS and MT' autonomous decision via LTE.
\section{Protocols and Procedures}\label{sec:protocols-procedures}
Here we present the technical concepts at the heart of TLTA, namely integration with the
handover at $sp$, operation in $pz$, and release at $op$.
\subsection{Trust-enhanced handover at the perimeter}\label{sec:trust-enhanc-hand}
\begin{figure}
\centerline{\subfigure{\resizebox{0.4\textwidth}{!}{\includegraphics{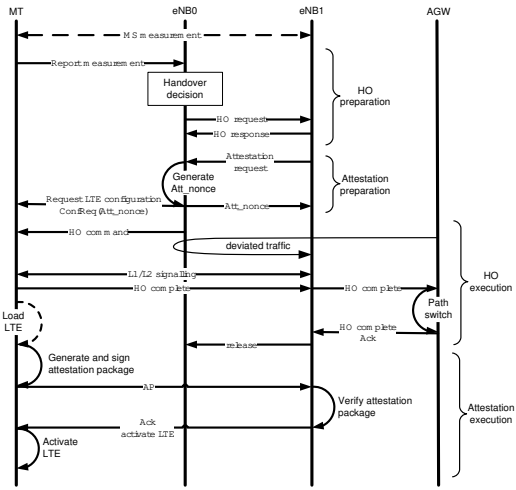}}}} 
\caption{Handover and attestation at the surveillance perimeter.}
  \label{fig:perimeter_handover}
\end{figure}

The trusted activation of the LTE\ie the activation of the LTE on the MT and a trustworthy 
report of that fact toward the network, when the device crosses the $sp$, is at the heart
of the TLTA concept.
The main idea is to integrate this process into the normal handover of MT between a cell
of $c_0$ and of  $c_1$.
Various handover protocols have been specified, and here we rely on 3GPPLTE, 
which offers advanced flexibility.
3GPPLTE handover (HO) is divided in two basic stages, HO preparation and HO execution.
In the present case it involves an eNB of $c_0$ (eNB0) and of $c_1$ (eNB1), as shown in 
Figure~\ref{fig:perimeter_handover}.
This process is augmented by two stages of remote attestation of the MT, regarding the presence and
activity of the LTE, in our scheme.
This design maximally separates the tasks between the eNBs of $c_0$ and $c_1$.
The first one, attestation preparation, sets up eNB1 and MT to perform LTE activation and remote
attestation.
The second one, attestation execution, is performed after HO and does not interfere with this 
time-critical process.

When eNB1 is notified of an HO request of some MT, it issues an HO response to eNB0, and additionally
notifies eNB0 of the necessity of a remote attestation of MT.
The sole task of eNB0 is then to generate and distribute a particular nonce, as
specified by the TCG's remote attestation protocol~\cite{RA}, for this purpose to 
MT and eNB1.
This is a modification of the original TCG attestation protocol which combines
eNB0 and eNB1 into one verifying party.
eNB0 issues the [TCG conforming] attestation request to MT, along with the 
\texttt{Att\_nonce}.
The activation of LTE and attestation thereof is concluded only after HO.
This means that the device loads the LTE (if it is not already present in 
its memory), generates
an attestation package\cite{RA},
and sends it to eNB1 only after the initial signalling with this B-Node
is complete and acknowledged by the MT.

By attestation, eNB1 can be convinced that MT is in a trustworthy state and has an
untampered LTE present.
It then sends an acknowledgement to MT, including the signal to activate the LTE
for operation within the $sp$.
\subsection{Protected zone operation and release}
\begin{figure}
\centerline{\subfigure{\resizebox{0.4\textwidth}{!}{\includegraphics{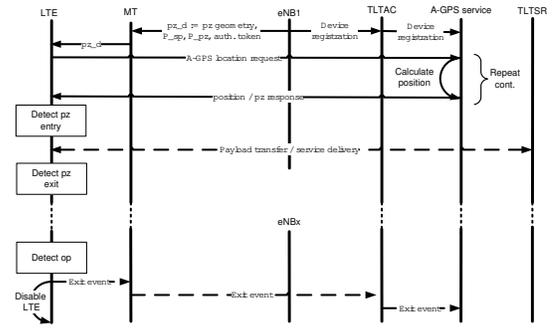}}}} 
\caption{Process after the activation of LTE and its interaction with TLTAC}
  \label{fig:TLTA_LTE_usage}
\end{figure}

Registration of the device at the TLTAC and download of the policy to the LTE on the MT
are shown in Figure~\ref{fig:TLTA_LTE_usage} and can be initiated by the AGW, eNB1 or the 
device itself. We have chosen the second variant in the figure.
After registration with TLTAC, and optionally the A-GPS service,
the MT/TLTAC pair can operate as a combination of 
policy enforcement point/policy decision point of an authorisation system in any desired 
separation of duties.

Policy enforcement can be divided into two levels for the area between  $sp$ and $pz$
and the interior of $pz$ with two separate policies called 
$\text{P}_{\textit{sp}}$ and $\text{P}_{\textit{pz}}$, respectively. 
Authorisation decisions can be felled either by LTE alone or by LTE in 
collaboration with LTACC in both cases. 
For the latter, LTE would transmit with an authorisation request a binary attribute to 
LTACC, stating whether the device is in \textit{pz} or not.

The location trigger is operated autonomously by the LTE, which periodically performs
GPS measurements and requests geopositioning from the A-GPS service.
When the MT detects entry to $pz$, the enforced policy is switched from 
$\text{P}_{\textit{sp}}$ to $\text{P}_{\textit{pz}}$.
MT can then, for instance, consume services from TLTSR.
Upon MT crossing $op$, it notifies TLTAC, which de-registers it and signals 
the exit event to the A-GPS service (optionally, also TLTSR could be signalled).
\section{Discussion}\label{sec:discussion}
In this section we discuss the security of TLTA, questions of performance
and practical issues, and finally sketch some concrete application scenarios.
\subsection{Security}
The two-tiered architecture of TLTA naturally divides security in two levels.
Security on the \textit{sp} rests on the integrity
of the involved B-nodes eNB0 and eNB1, which is a precondition. 
Security in \textit{pz} on the other hand depends on the separation of
tasks between TLTAC and the LTE. 
The security of the LTE is enabled by the unique features of
trust-enhanced MTs, namely secure boot and attestation, features which are
fundamental for TLTA.
When a registered device enters or leaves 
pz from or to sp, policy enforcement by LTE switches between the two policies 
$\text{P}_{\textit{sp}}$ and $\text{P}_{\textit{pz}}$ above.
We discuss the security properties of the enforcement of $\text{P}_{\textit{pz}}$.

The design entails that a security assessment of TLTA must distinguish between
policies for \textit{functional enforcement} and \textit{access control}, respectively.
The former means that certain functions on \textit{all devices in pz} shall be enabled or
disabled. 
The latter allows certain content or services to be delivered or made
available to the MTs registered with TLTAC. 
The essential difference between the two is that, in order to extend functional
enforcement to all MTs in \textit{pz}, LTE activation must be ensured on all of them or
the policy must be enforced otherwise\eg by out-of-band measures and processes.
Access control is less security-critical.
In contrast to functional enforcement, 
an access control policy is effectively enforced by
the TLTAC, which will yield access only to MTs with active LTE accepting $\text{P}_{\textit{pz}}$. 
In this case, LTE provides the added security that proliferation of access
to content or services beyond \textit{pz} is effectively prevented.

Functional enforcement in TLTA is susceptible to certain attacks which do not apply
to access control and whose general strategy is to subvert \textit{sp}.
The simplest and most effective is to cross \textit{sp} and enter \textit{pz} with a disabled
MT, or an MT shielded against network communication.
Though one could attempt to detect such rogue MTs via the network when they are activated 
inside \textit{pz}, this might not be very efficient nor effective.
A more sophisticated attacker could try to prevent the handover at the sp\eg using 
a beam antenna to tamper the MT's measurement of the base stations.
In principle, this kind of attacks can only be mitigated by additional, physical access
control procedures on \textit{pz}'s boundary.

A general caveat of architectures relying on a trusted platform built upon a 
hardware security anchor like an MTM is that it is more susceptible
to side channel attacks than\eg a smart card.
Applied to  the TLTA concept this argument could mean to imply that\eg location data
is tampered with while in the MT's memory or in transit between processor and memory,
respectively, the wireless modem.
A hardening of the MT preventing this could be achieved using secure 
channels~\cite{WIS07,ATMEL,YGZ05} (Texas Instruments' proprietary M-Shield technology
is another example which could sensibly be combined with Trusted Computing
for TLTA and other purposes).
\subsection{Efficiency}
The activities during the central stage of TLTA, the entrance to the surveillance perimeter,
offer a great variety of implementation options.
In particular it should be noted that though we used (A)-GPS as a prime
example, any other autonomous or MT/eNB co-operative  method for
detecting entry to \textit{pz} such as triangulation, is applicable.  

Let us briefly discuss the design choices taken in 
Sections~\ref{sec:tlta-concept} and~\ref{sec:protocols-procedures}.
The de-centralised architecture and operation of TLTA is designed with
the central aim of reducing network load.
This is only possibly with the desired level of security and enforcement
if trust-enhanced MTs are used as central functional building blocks.
TLTA is loosely coupled to the top-level of the cellular infrastructure.
The TLTAC operates on the level of the network's Home/Visitor Location Register
(HLR/VLR) and Mobile Switching Centre (MSC) And could be co-located or
integrated with either one. 

Signalling cost is a major concern for location management
in cellular overlay networks~\cite{AW02,MA07}, which poses similar demands
as location based authorisation. 
In our case, the two-tiered method of TLTA yields the largest advantage 
in this respect by leveraging the capacities of the edge of the network.
In particular communication with MTs at the \textit{sp} is completely borne by
the eNBs and thus maximally de-centralised, and moreover
MTs operate largely autonomous inside \textit{pz}.
Furthermore, the protocol is designed such that the eNB of $c_0$ is passive, its 
sole essential task being the generation of the \verb+Att_nonce+.
This saves communication effort at the stage of node configuration, since
only the eNBs of $c_1$ need to be activated to handle entry events. 
Another feature is the separation of inbound (\textit{sp}) and outbound
perimeter. It avoids judder\ie frequent switching on and off of the LTE
and according (de)registration with the TLTAC.

A bottleneck could ensue if the physical access to the \textit{pz} is restricted,
in the extreme case to a single cell and a large number of MTs enters \textit{sp}.
This can happen in particular in application scenarios for functional enforcement
(see below). This concentrates the load for the TLTA entry protocol to
a single eNB0, eNB1 pair.
The problem can be alleviated by a) organisational measures to distribute
entry points over the perimeter of \textit{pz}, or b) shifitng the \textit{sp}
further away from \textit{pz}\ie to the boundary between cell layers $c_{-n}$
$c_{-n+1}$, $n\geq1$. A third option is to introduce extra cells \textit{ad hoc}\eg
mobile nano/pico-cells\cite{HS00,Mol00,Yam05}. 
This is also an interesting option if the network's 
deployed B-nodes in the area are not capable to execute the TLTA protocol.

Regarding the mobile users respectively device owners, 
the forceful activation of the LTE is
a potential infringement of the proprietor's power of disposal.
A similar consideration holds for Trusted Computing technology proper,
on any platform. It has led the TCG to mandate activation and take ownership
procedures for the hardware security anchor TPM/MTM.
These concepts can sensibly be lifted to the level of applications
intimately connected with and depending on TC.
Furthermore, procedures for user notification and acknowledgement
should be combined with TLTA.
\subsection{Application Scenarios}
The TLTA technology is of basic character, hence it is easy to envisage
application scenarios in many sectors ranging from general m-commerce to
public safety. We note but a few.
\subsubsection{High-tech trade fair}
Functional restriction policies can be enforced with exemptions. 
Consider a high-tech trade fair. \textit{pz} is determined by the fair's 
host so as to coincide with the physical entry barriers of the fair area. 
\textit{sp} is a larger area so that visitors with TLTAC-enabled MTs
will already be registered with TLTAC when passing the entry control.
$\text{P}_{\textit{pz}}$
stipulates that cameras and sound recording facilities 
of devices are to be switched off by LTE.   
They then can freely pass through, while unprotected 
(or non-networked devices\eg cameras)  are detected at the 
entry barrier and are confiscated.  
Countering industrial espionage is another use case.
A company may use TLTA to disable MT cameras within the borders of their campus,
while keeping the MTs of visiting executives operative for communication purposes.
\subsubsection{Sports stadium and concert hall}
The voice commentary on a sports event is broadcast 
encrypted to mobile devices in a stadium, 
which is the \textit{pz}. 
Only TLTA-enabled devices receive the decryption keys
with the $\text{P}_{\textit{pz}}$ data managed securely by LTE.
The TLTSR and service provider can simply broadcast the content
and has no need to operate own access control facilities.
In a concert hall, the live audio broadcast should add to the experience of the audience, 
and can be personalised in terms of language and otherwise. 
This shall not be able to be received outside of the concert hall, to avoid bootlegging.
\subsubsection{Mission critical communication}
MTs in a disaster area can be configured to receive emergency and alarming messages.
This can be used to inform the public or emergency action forces but also to
locate them within the danger zone. Standard mobile devices can co-operate
with the special ones of action forces and inter-work over network boundaries\eg
with TETRA networks~\cite{Sal06,Ed06}.
\subsubsection{Mobile gaming}
A game spread out over the area of a city requires to adjust the 
device side policies based on narrow location\eg when the user has reached a
certain waypoint.

These scenarios have in common that they could in principle also be realised 
by authorisation not based on location\eg organisational processes, yet with
varying methods to reach the desired enforcement level.
The benefit of TLTA is a unified authorisation method scaling with security
requirements.
\section{Conclusion}\label{sec:conclusion}
We have presented a novel system for the enforcement of authorisation
policies according to location attributes.
The salient features of i) largely de-central operation, and ii) strong enforcement
level rest on a few key ideas.
First, trust-enhanced devices combine functions of secure location triggering with
with policy enforcement.
Second, establishment of the protected zone policy enforcement is efficiently
integrated in the handover between cells.
Third, location trigger operation is two-tiered combining cell-based location
with precision location methods like (A)-GPS.
These concepts optimally leverage network and device resources and concurrently 
provide a high security level.
Potential applications abound, some of which are only possible with
TLTA or related concepts employing trusted devices.
Implementation of TLTA seems feasible and economically viable, once 
TC-enabled devices become widely available.







\begin{thebibliography}{27}
\providecommand{\url}[1]{#1}
\csname url@rmstyle\endcsname
\providecommand{\newblock}{\relax}
\providecommand{\bibinfo}[2]{#2}
\providecommand\BIBentrySTDinterwordspacing{\spaceskip=0pt\relax}
\providecommand\BIBentryALTinterwordstretchfactor{4}
\providecommand\BIBentryALTinterwordspacing{\spaceskip=\fontdimen2\font plus
\BIBentryALTinterwordstretchfactor\fontdimen3\font minus
  \fontdimen4\font\relax}
\providecommand\BIBforeignlanguage[2]{{%
\expandafter\ifx\csname l@#1\endcsname\relax
\typeout{** WARNING: IEEEtran.bst: No hyphenation pattern has been}%
\typeout{** loaded for the language `#1'. Using the pattern for}%
\typeout{** the default language instead.}%
\else
\language=\csname l@#1\endcsname
\fi
#2}}

\bibitem{Anis07}
M. Anisetti, C.A. Ardagna, V. Bellandi, E. Damiani: \emph{OpenAmbient: a Pervasive Access Control Architecture}.
In: Long-Term and Dynamical Aspects of Information Security: Emerging Trends in Information and Communication Security,
A. U. Schmidt, M. Kreutzer, R. Accorsi (eds.), Nova Science Publishers, Hauppauge, NY, 2007.


\bibitem{Woel02} 
G. W\"olfle, R. Hoppeand, D. Zimmermann, F. M. Landstorfer, 
\emph{Enhanced Localization Technique within Urban and Indoor Environments based on Accurate and Fast
Propagation Models}. European Wireless 2002, Firence, Italy, February 2002.

\bibitem{WHZ+02}
P. Wertz, R. Hoppe, D. Zimmermann, G. W\"olfle, F. M. Landstorfer,
\emph{Enhanced Localization Technique within Urban and Indoor Environments}. 
3rd COST 237 MCMMeeting, Guildford, UK, January 2002.

\bibitem{VNJ02}
J. Vidal, J., M. Najar, R. Jativa, 
\emph{High resolution time-of-arrival detection for wireless positioning systems}. 
In: Proc. of the 56th IEEE  Vehicular Technology Conference, 2002.

\bibitem{MPWGRefArch}
Trusted Computing Group, \emph{TCG Mobile Reference Architecture}.
Specification version 1.0, Revision 1. 12 June 2007.
\url{https://www.trustedcomputinggroup.org/specs/mobilephone/}

\bibitem{3GPPLTE}
H. Ekstrom, A. Furuskar, J. Karlsson, M. Meyer, S. Parkvall, J. Torsner, M. Wahlqvist, 
\emph{Technical solutions for the 3G long-term evolution}, 
IEEE Communications Magazine \textbf{44}, no.3, 38--45, March 2006.

\bibitem{RA}
Trusted Computing Group, \emph{TPM Main Part 2 TPM Structures}, 
Specification Version 1.2 Revision 94, March 2006.
\emph{TPM Main Part 3 Commands}, Specification Version 1.2 Revision 94, March 2006.

\bibitem{WIS07} 
A. P. Wise, \emph{Memory encryption for digital video}. U.S. Patent US2007140477 (A1). 2007-06-20.

\bibitem{ATMEL}
Data sheets from the ATMEL Corporation.
\url{http://www.atmel.com/dyn/products/datasheets.asp?family_id=646}

\bibitem{YGZ05}
J. Yang, L.  Gao, Y. Zhang,  \emph{Improving Memory Encryption Performance in Secure Processors}. 
IEEE Transactions on Computers  \textbf{54}  (2005)  630 - 640.

\bibitem{AW02}
I. F. Akyildiz, W. Wang, \emph{A dynamic location management scheme for next-generation multitier PCS systems}. 
IEEE Transactions on Wireless Communications \textbf{1}, no.1, 178--189, January 2002.

\bibitem{MA07}
D. Morris, A. H. Aghvami, \emph{Location Management Strategies for Cellular 
Overlay Networks --- A Signaling Cost Analysis}.
IEEE Transactions on Broadcasting \textbf{53}, no.2, 480--493, June 2007.


\bibitem{HS00}
C. Hartmann, O. Schlegelmilch, \emph{Hierarchical cell structures with adaptive radio resource management}. 
52nd IEEE Vehicular Technology Conference, 2000. VTS-Fall VTC 2000.  Vol.4, pp. 1764--1771, 2000.

\bibitem{Mol00}
D. Molkdar, \emph{Simulation results of a typical GSM picocellular system}. 
52nd IEEE Vehicular Technology Conference, 2000. VTS-Fall VTC 2000.  Vol.4, pp. 1590--1596, 2000.

\bibitem{Yam05}
T. Yamada, \emph{Another double tier to avoid local congestion in a unified microcellular network}.
IEEE 16th International Symposium on Personal, Indoor and Mobile Radio Communications, 2005. 
PIMRC 2005. Vol.4, pp. 2232--2236, 11-14 Sept. 2005

\bibitem{Sal06}
 A. K. Salkintzis, \emph{Evolving public safety communication systems by integrating WLAN and TETRA networks},
IEEE Communications Magazine \textbf{44}, no.1, 38--46, Jan. 2006.

\bibitem{Ed06}
C. Edwards, \emph{Wireless - Building on Tetra}, 
Engineering \& Technology \textbf{1}, no.2, 32--36, May 2006.

\end{thebibliography}
%


\providecommand{\noopsort}[1]{} \providecommand{\singleletter}[1]{#1}

\end{document}